\documentclass[a4]{jpconf}
\usepackage{citesort}
\usepackage[dvipdfmx]{graphicx}
\usepackage{amsmath,amssymb}
\begin{document}
\title{Quantum spin liquid and electric quadrupolar states 
of single crystal Tb$_{2+x}$Ti$_{2-x}$O$_{7+y}$}

\author{M Wakita, T Taniguchi, H Edamoto, H Takatsu and H Kadowaki}
\address{Department of Physics, Tokyo Metropolitan University, Hachioji-shi, Tokyo 192-0397, Japan}

\begin{abstract}
The ground states of the frustrated pyrochlore oxide 
Tb$_{2+x}$Ti$_{2-x}$O$_{7+y}$, sensitively depending on the 
small off-stoichiometry parameter $x$, have been studied 
by specific heat measurements using well characterized samples. 
Single crystal Tb$_{2+x}$Ti$_{2-x}$O$_{7+y}$ boules grown 
by the standard floating zone technique are shown to 
exhibit concentration ($x$) gradient.
This off-stoichiometry parameter is determined by 
precisely measuring the lattice constant of small samples 
cut from a crystal boule. 
Specific heat shows that the phase boundary of the electric 
quadrupolar state has a dome structure 
in the $x$-$T$ phase diagram with the highest 
$T_{\text{c}} \simeq 0.5$ K at about $x = 0.01$. 
This phase diagram suggests that the putative 
U(1) quantum spin-liquid state of Tb$_{2+x}$Ti$_{2-x}$O$_{7+y}$ exists 
in the range $x<x_{\text{c}} \simeq -0.0025 $, 
which is separated from the quadrupolar state 
via a first-order phase-transition line $x=x_{\text{c}}$. 
\end{abstract}

\section{Introduction}
Magnetic systems with geometric frustration have been intensively studied 
experimentally and theoretically for decades \cite{Lacroix11}. 
Spin systems on networks of triangles or tetrahedra, 
such as triangular \cite{Wannier50}, kagom\'{e} \cite{Shyozi51}, 
and pyrochlore \cite{Gardner10} lattices, play major roles 
in these studies. 
Subjects fascinating many investigators in recent years are quantum 
spin liquid (QSL) states \cite{Lee08,Balents10}, 
where conventional long-range orders (LRO) are suppressed 
to very low temperatures. 

Among frustrated magnetic pyrochlore oxides \cite{Gardner10}, 
Tb$_2$Ti$_2$O$_7$ (TTO) has attracted much attention 
because it does not show any conventional LRO down to 50 mK \cite{Gardner99}, 
suggesting that it is a candidate for a QSL state. 
Although many experimental studies of TTO have been performed 
to date, the problem why TTO does not show any magnetic LRO 
remains very difficult \cite{Gingras14,Petit15}. 
This is partly because TTO shows strong sample dependence \cite{Takatsu12}, 
extremely strong for single crystals. 
And accordingly, simple interpretation of experimental data 
is precluded. 

Recently, we investigated polycrystalline samples of 
off-stoichiometric Tb$_{2+x}$Ti$_{2-x}$O$_{7+y}$, 
and showed that a very small change of $x$ induces a quantum phase 
transition between a spin liquid state ($x < -0.0025 = x_{\text{c}} $) 
and a LRO state with a hidden order parameter ($x_{\text{c}} < x$) \cite{Taniguchi13}. 
The $x$-$T$ phase diagram of Tb$_{2+x}$Ti$_{2-x}$O$_{7+y}$ suggested 
in Ref. \cite{Taniguchi13} has a dome-shape LRO phase boundary. 
More recently, we study the hidden LRO using an $x$-controlled 
single crystal, which shows 
a very sharp peak in specific heat at $T_{\text c} = 0.53$ K ($ x \simeq 0.005 $) \cite{Takatsu15}. 
By using semi-quantitative analyses, we propose \cite{Takatsu15,Kadowaki15,Takatsu15B} 
that the LRO of Tb$_{2+x}$Ti$_{2-x}$O$_{7+y}$ is an electric multipolar 
(or quadrupolar) state. 
This LRO state was theoretically predicted \cite{Onoda11} 
using electronic superexchange interactions for non-Kramers ions, 
including Tb$^{3+}$, 
which have both magnetic dipole and electric quadrupole (16-pole, and 64-pole) moments. 
In addition, quite intriguingly, the estimated parameter set \cite{Takatsu15} 
of the effective pseudospin-1/2 Hamiltonian is located 
very close to a theoretical phase boundary between the electric quadrupolar 
and U(1) quantum spin-liquid states \cite{Onoda11,Lee12}, 
which could naturally explain the spin liquid state of TTO. 

The purpose of this investigation is to extend 
our study of polycrystalline Tb$_{2+x}$Ti$_{2-x}$O$_{7+y}$ \cite{Taniguchi13} 
to single crystals in the hope that the above scenario 
for the TTO problem is reinforced. 
We grow single crystals of Tb$_{2+x}$Ti$_{2-x}$O$_{7+y}$ 
by the standard floating zone (FZ) technique \cite{Gardner98} 
and have found that very precise measurements of the lattice constant 
are useful to characterize the single crystals. 
Specific heat of these samples with different off-stoichiometry 
parameters ($x$) have been measured down to 0.1 K to obtain 
an $x$-$T$ phase diagram.

\section{Experimental methods and results}
Polycrystalline samples of Tb$_{2+x}$Ti$_{2-x}$O$_{7+y}$ were 
prepared by the standard solid-state reaction 
as described in Ref. \cite{Taniguchi13}. 
The two starting materials, Tb$_{4}$O$_{7}$ and TiO$_2$, 
were heated in air at 1350~$^\circ$C for several days with 
periodic grindings to ensure a complete reaction. 
The value of $x$ was adjusted by changing the mass ratio 
of the two materials, and is nominal with an offset 
about $\pm 0.002$. 
The resulting Tb$_{2+x}$Ti$_{2-x}$O$_{7+y}$ powder samples 
were used for single crystal growth 
by the standard FZ technique \cite{Gardner98}. 
Crystal growth was carried out in an Ar gas flow atmosphere 
using a double ellipsoidal image furnace (NEC SC-N35HD).  

X-ray powder-diffraction experiments were carried out 
using a RIGAKU-SmartLab diffractometer 
equipped with a Cu K$_{\alpha 1}$ monochromator. 
To precisely measure the lattice constant 
we performed $\theta$-$2\theta$ scans on powder mixtures 
of polycrystalline or crushed-crystalline Tb$_{2+x}$Ti$_{2-x}$O$_{7+y}$ 
and Si \cite{Taniguchi13,Taniguchi15}. 
Absolute values of lattice constants are normalized by using 
the certified lattice parameter for a temperature of 22.5 $^\circ$C 
of the SRM-640d Si powder, $a=5.43123$ {\AA} \cite{SRM640d}, 
being further corrected for the temperature dependence \cite{Izumi09}. 

\begin{figure}[b]
\begin{center}
\includegraphics[width=16.0cm,clip]{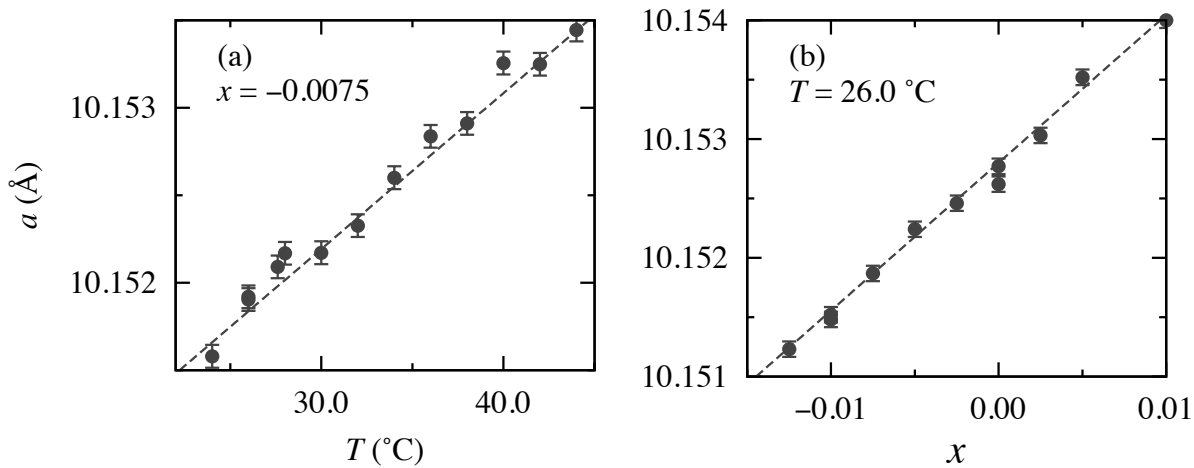}
\end{center}
\caption{\label{TdepLC}
Lattice constant $a(T,x)$ of polycrystalline Tb$_{2+x}$Ti$_{2-x}$O$_{7+y}$ samples.
(a) Temperature dependence of $a(T,x=-0.0075)$.
(b) Off-stoichiometry parameter dependence of $a(T=26.0^{\circ} \text{C},x)=0.124418x+10.15280$ from Ref. \cite{Taniguchi15}.
}
\end{figure}

Temperature dependence of 
the lattice constant $a(T,x)$ of Tb$_{2+x}$Ti$_{2-x}$O$_{7+y}$ 
was measured using a polycrystalline 
sample with $x=-0.0075$, and the result is shown in Fig.~\ref{TdepLC}(a). 
The $x$ dependence of $a(T=26.0^{\circ} \text{C},x)$ of polycrystalline 
samples is plotted in Fig.~\ref{TdepLC}(b), where 
we converted the published lattice constants (Fig. 1 in Ref.~\cite{Taniguchi13}) 
to those at 26.0~$^{\circ}$C \cite{Taniguchi15}. 

\begin{figure}
\begin{center}
\includegraphics[width=12.0cm,clip]{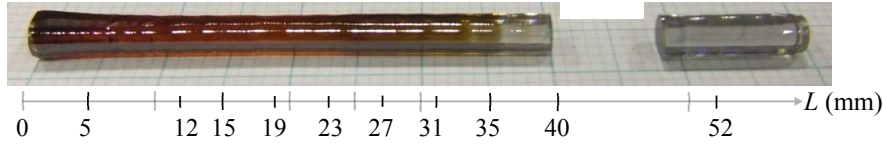}
\end{center}
\caption{\label{crystal_boule}
Single crystal Tb$_{2+x}$Ti$_{2-x}$O$_{7+y}$ boule grown by the FZ method, 
where the missing part ($40<L<48$ mm) was cut before taking this photograph. 
The numbers represent distances $L$ along the growth direction, where 
small crystals are cut at these $L$ values. 
Lattice constant and specific heat of these crystals are shown 
in Figs.~\ref{crystal_LC} and \ref{specific_heat}.
}
\end{figure}

\begin{figure}
\begin{center}
\includegraphics[width=12.0cm,clip]{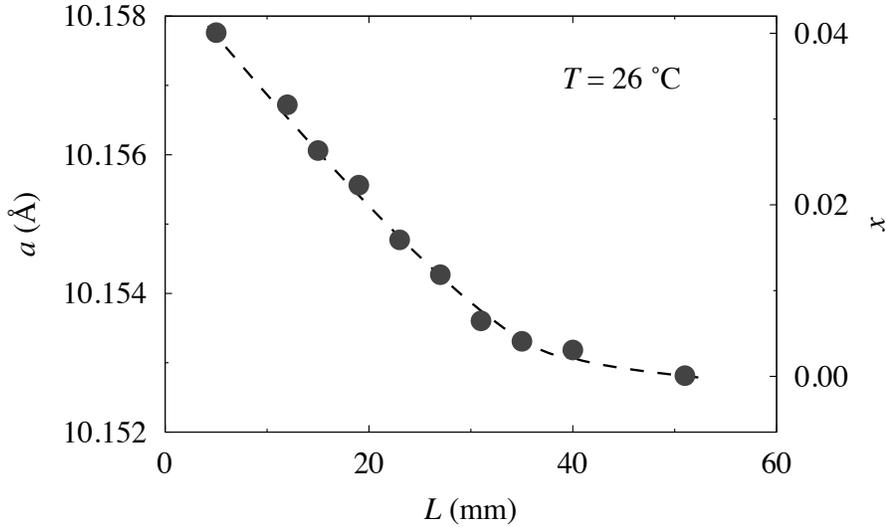}
\end{center}
\caption{\label{crystal_LC}
Lattice constants of small Tb$_{2+x}$Ti$_{2-x}$O$_{7+y}$ crystals 
cut from the boule shown in Fig.~\ref{crystal_boule}. 
These lattice constants are converted to $x$ using 
the polycrystalline curve, i.e., $a(T=26.0^{\circ} \text{C},x)$ of Fig.~\ref{TdepLC}(b) 
and are shown on the right vertical axis. 
}
\end{figure}

Figure \ref{crystal_boule} shows a single crystal Tb$_{2+x}$Ti$_{2-x}$O$_{7+y}$ boule 
that was grown from a feed rod of $x=-0.005$ powder and was post-annealed for about 7 days 
at 1000~$^{\circ}$C in air. 
Lattice constants of small crystals cut from this boule 
were measured at 26.0~$^{\circ}$C and are plotted as a function of 
the distance along the growth direction $L$ shown in Fig.~\ref{crystal_LC}. 
We assume that $a(T=26.0^{\circ} \text{C},x)$ of polycrystalline samples 
(Fig.~\ref{TdepLC}(b)) and its linear extension to the range $x>0.01$ 
can be used to estimate the off-stoichiometry parameter ($x$) of the small crystals. 
These $x$ values are shown on the right vertical axis of Fig.~\ref{crystal_LC}. 
One can see that the boule has a systematic $x$ gradient. 
During the crystal growth the off-stoichiometry parameter starts 
from $x \simeq 0.04$ ($L =$ 1 -- 5 mm), 
then decreases linearly as a function of $L$, 
and finally varies more slowly ($L > 40$ mm). 

\begin{figure}
\begin{center}
\includegraphics[width=16.0cm,clip]{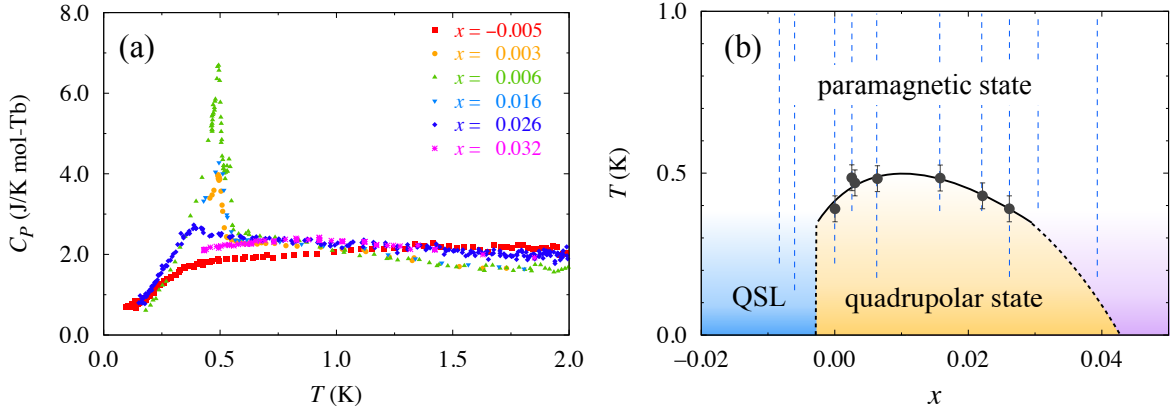}
\end{center}
\caption{\label{specific_heat}
(a) Temperature dependence of specific heat of several single crystals. 
The $x$ values are estimated by the method shown in Fig.~\ref{crystal_LC}. 
(b) $x$-$T$ phase diagram determined from the specific heat measurements of single crystals. 
Temperature ranges of the specific heat measurements are shown by vertical blue dashed lines. 
}
\end{figure}

To characterize crystal samples we also measured specific heat $C_P(T)$ 
at low temperatures using a $^{3}$He or 
an adiabatic demagnetization refrigerator. 
In Fig.~\ref{specific_heat}(a) we show specific heat as a function of temperature 
for the several crystals cut from the boule (Fig.~\ref{crystal_boule}) 
and a few from another boule. 
Based on these $C_P(T)$ data we draw a tentative $x$-$T$ phase diagram for the single crystals 
in Fig.~\ref{specific_heat}(b). 
We note that these $C_P(T)$ data and the $x$-$T$ phase diagram for 
the single crystals are quite consistent with 
those of polycrystalline Tb$_{2+x}$Ti$_{2-x}$O$_{7+y}$ \cite{Taniguchi13}. 
This indicates that our trial method of estimating small $x$ ($|x| < 0.01$)
for single crystals 
using the precise measurement of the lattice constant is probably reliable. 

The $x$-$T$ phase diagram (Fig.~\ref{specific_heat}(b)) implies that 
one has to take a special care of very small change 
of the off-stoichiometry existing even in a single crystal boule 
to investigate Tb$_{2+x}$Ti$_{2-x}$O$_{7+y}$ (or nominal Tb$_{2}$Ti$_{2}$O$_{7}$). 
Previous experimental investigations using small TTO crystals 
will have to be reinterpreted as investigations on different 
Tb$_{2+x}$Ti$_{2-x}$O$_{7+y}$ crystals. 
In particular, previous experiments using large crystals, 
especially inelastic neutron scattering for example 
Refs. \cite{Yasui02,Fennel14,Guitteny13,Fritsch13}, 
require special caution in their interpretation, 
because the crystals may not be sufficiently homogeneous.

\section{Discussion and summary}
The $x$-$T$ phase diagram shows that 
around $x=x_{\text{c}} \simeq -0.0025$ the transition temperature $T_{\text{c}}$ 
of the quadrupolar state \cite{Takatsu15} disappears abruptly in a small $x$ range. 
This suggests that the neighboring putative QSL state  
is separated by a first-order phase-transition line $x = x_{\text{c}}$ \cite{Takatsu15,Taniguchi13}. 
It is interesting that this type of first-order phase transition 
between U(1) QSL and quadrupolar states is predicted 
by a gauge mean-field theory \cite{Lee12}, presumably relevant to TTO \cite{Takatsu15}. 
One may naturally expect that 
Tb$_{2+x}$Ti$_{2-x}$O$_{7+y}$ with $x=x_{\text{c}}$ 
is on the theoretical border of U(1) QSL and quadrupolar states \cite{Lee12}, 
and that the spin liquid state of Tb$_{2+x}$Ti$_{2-x}$O$_{7+y}$ with $x<x_{\text{c}}$ 
is U(1) QSL of Ref. \cite{Lee12}. 
This is a very intriguing hypothesis for further studies. 

On the other hand, in a larger $x$ range of $x>0.01$ the transition temperature 
of the quadrupolar state seems to decrease gradually 
and the specific heat peak gradually becomes smaller as $x$ is increased. 
These suggest that an effect of randomness controls the system. 
A possible scenario of the randomness effect may be as follows. 
Most of excess Tb atoms reside on the Ti$^{4+}$ site and become Tb$^{4+}$ ions. 
These magnetic Tb$^{4+}$ ions behave as magnetic impurities in the system, 
where local magnetic short-range order is restored around each Tb$^{4+}$ ion. 
The quadrupolar state is completely suppressed in $x>0.04$.

In summary, we have investigated single-crystalline samples of 
the frustrated pyrochlore oxide 
Tb$_{2+x}$Ti$_{2-x}$O$_{7+y}$ 
by growing single crystals using the standard floating zone technique 
and by characterizing them using X-ray diffraction techniques 
and specific heat measurements down to 0.1 K. 
We show that a precise determination of the lattice constant 
is useful for estimating the small off-stoichiometry parameter $x$. 
Small crystals cut from single crystal rods, exhibiting $x$ gradient, 
show three different low temperature behaviors:
a paramagnetic QSL ($x<x_{\text{c}}$), 
a long range quadrupolar, and 
possibly a randomness dominating state. 
The phase boundary of the quadrupolar state shows 
a dome structure in the $x$-$T$ phase diagram with the highest 
$T_{\text{c}} \simeq 0.5$ K at $x=0.01$ and suggests 
existence of a first-order phase-transition line 
separating the QSL and quadrupolar states.

\ack 
We thank S. Onoda and Y. Kato 
for useful discussions. 
This work was supported by JSPS KAKENHI grant numbers 25400345 and 26400336.

\section*{References}
\bibliography{TTO_MW}

\providecommand{\newblock}{}
\begin{thebibliography}{10}
\expandafter\ifx\csname url\endcsname\relax
  \def\url#1{{\tt #1}}\fi
\expandafter\ifx\csname urlprefix\endcsname\relax\def\urlprefix{URL }\fi
\providecommand{\eprint}[2][]{\url{#2}}

\bibitem{Lacroix11}
Lacroix C, Mendels P and Mila F (eds) 2011 {\em Introduction to Frustrated
  Magnetism\/} (Berlin, Heidelberg: Springer)

\bibitem{Wannier50}
Wannier G~H 1950 {\em Phys. Rev.\/} {\bf 79} 357

\bibitem{Shyozi51}
Sy\^ozi I 1951 {\em Prog. Theor. Phys.\/}  306

\bibitem{Gardner10}
Gardner J~S, Gingras M~J~P and Greedan J~E 2010 {\em Rev. Mod. Phys.\/} {\bf
  82} 53

\bibitem{Lee08}
Lee P~A 2008 {\em Science\/} {\bf 321} 1306

\bibitem{Balents10}
{Balents} L 2010 {\em Nature\/} {\bf 464} 199

\bibitem{Gardner99}
Gardner J~S, Dunsiger S~R, Gaulin B~D, Gingras M~J~P, Greedan J~E, Kiefl R~F,
  Lumsden M~D, MacFarlane W~A, Raju N~P, Sonier J~E, Swainson I and Tun Z 1999
  {\em Phys. Rev. Lett.\/} {\bf 82} 1012

\bibitem{Gingras14}
Gingras M~J~P and McClarty P~A 2014 {\em Rep. Prog. Phys.\/} {\bf 77} 056501

\bibitem{Petit15}
Petit S, Guitteny S, Robert J, Bonville P, Decorse C, Ollivier J, Mutka H and
  Mirebeau I 2015 {\em EPJ Web of Conferences\/} {\bf 83} 03012

\bibitem{Takatsu12}
{Takatsu} H, {Kadowaki} H, {Sato} T~J, {Lynn} J~W, {Tabata} Y, {Yamazaki} T and
  {Matsuhira} K 2012 {\em J. Phys. Condens. Matter\/} {\bf 24} 052201

\bibitem{Taniguchi13}
Taniguchi T, Kadowaki H, Takatsu H, F\aa{}k B, Ollivier J, Yamazaki T, Sato
  T~J, Yoshizawa H, Shimura Y, Sakakibara T, Hong T, Goto K, Yaraskavitch L~R
  and Kycia J~B 2013 {\em Phys. Rev. B\/} {\bf 87} 060408

\bibitem{Takatsu15}
Takatsu H, Kittaka S, Kasahara A, Kono Y, Sakakibara T, Kato Y, Onoda S,
  F\aa{}k B, Ollivier J, Lynn J W, Taniguchi T, Wakita M and Kadowaki H
  arXiv:1506.04545

\bibitem{Kadowaki15}
{Kadowaki} H, {Takatsu} H, {Taniguchi} T, {F{\aa}k} B and {Ollivier} J 2015
  {\em Spin\/} {\bf 5} 1540003

\bibitem{Takatsu15B}
Takatsu H, Taniguchi T, Kittaka S, Sakakibara T and Kadowaki H, J Phys. CS 2016
  TMU meeting Sept. 2015

\bibitem{Onoda11}
Onoda S and Tanaka Y 2011 {\em Phys. Rev. B\/} {\bf 83} 094411

\bibitem{Lee12}
Lee S, Onoda S and Balents L 2012 {\em Phys. Rev. B\/} {\bf 86} 104412

\bibitem{Gardner98}
Gardner J, Gaulin B and Paul D 1998 {\em J. Crystal Growth\/} {\bf 191} 740

\bibitem{Taniguchi15}
Taniguchi T, Kadowaki H, Takatsu H, F\aa{}k B, Ollivier J, Yamazaki T, Sato
  T~J, Yoshizawa H, Shimura Y, Sakakibara T, Hong T, Goto K, Yaraskavitch L~R
  and Kycia J~B 2015 {\em Phys. Rev. B\/} {\bf 92} 019903

\bibitem{SRM640d}
Kaiser D~L and Watters R~L 2010 {}NIST Certificate of SRM 640d,\\
  \verb|https://www-s.nist.gov/srmors/view_cert.cfm?srm=640D|

\bibitem{Izumi09}
Izumi F and Nakai I (eds) 2009 {\em Funmatsu Xsen kaiseki no jissai\/} 2nd ed
  (Tokyo: Asakurashoten)

\bibitem{Yasui02}
{Yasui} Y, {Kanada} M, {Ito} M, {Harashina} H, {Sato} M, {Okumura} H, {Kakurai}
  K and {Kadowaki} H 2002 {\em J. Phys. Soc. Jpn.\/} {\bf 71} 599

\bibitem{Fennel14}
Fennell T, Kenzelmann M, Roessli B, Mutka H, Ollivier J, Ruminy M, Stuhr U,
  Zaharko O, Bovo L, Cervellino A, Haas M~K and Cava R~J 2014 {\em Phys. Rev.
  Lett.\/} {\bf 112} 017203

\bibitem{Guitteny13}
Guitteny S, Robert J, Bonville P, Ollivier J, Decorse C, Steffens P, Boehm M,
  Mutka H, Mirebeau I and Petit S 2013 {\em Phys. Rev. Lett.\/} {\bf 111}
  087201

\bibitem{Fritsch13}
Fritsch K, Ross K~A, Qiu Y, Copley J~R~D, Guidi T, Bewley R~I, Dabkowska H~A
  and Gaulin B~D 2013 {\em Phys. Rev. B\/} {\bf 87} 094410

\end{thebibliography}

\end{document}